# Detection of Fraudulent Sellers in Online Marketplaces using Support Vector Machine Approach

Shini Renjith

*Assistant Professor, Department of Computer Science & Engineering*
*Mar Baselios College of Engineering and Technology*
*Nalanchira, Paruthippara, Thiruvananthapuram, Kerala, India*

*shinirenjith@gmail.com*

**Abstract**-*The e-commerce share in the global retail spend is showing a steady increase over the years indicating an evident shift of consumer attention from bricks and mortar to clicks in retail sector. In recent years, online marketplaces have become one of the key contributors to this growth. As the business model matures, the number and types of frauds getting reported in the area is also growing on a daily basis. Fraudulent e-commerce buyers and their transactions are being studied in detail and multiple strategies to control and prevent them are discussed. Another area of fraud happening in marketplaces are on the seller side and is called merchant fraud. Goods/services offered and sold at cheap rates, but never shipped is a simple example of this type of fraud. This paper attempts to suggest a framework to detect such fraudulent sellers with the help of machine learning techniques. The model leverages the historic data from the marketplace and detect any possible fraudulent behaviours from sellers and alert to the marketplace.*

## I. INTRODUCTION

A recent report from eMarketer (www.emarketer.com), a leading market research company in the area of digital marketing and commerce, estimates the total international e-commerce sales volume to touch $4 trillion in 2020 when the total retail sale is estimated as $27 trillion. This is 14.6% of the total retail spend expected. This is a considerable change in the market share as in 2016 only 8.7% of the international retail sales are happening from e-commerce amounting to $1.915 trillion out of $22.049 trillion [Fig. 1]. There are multiple factors contributing to the growth of the sector and the first and foremost one is the change in consumer behaviour who likes to do compare and buy at the comfort of home/office. Retailers are also interested in the model as it proved cheaper than the traditional model for them and they are ready to share a portion of their profit to the consumer making products and services available at cheaper price at online. The next movement in e-commerce business model was the evolution of online marketplaces where multiple sellers share a common selling platform.

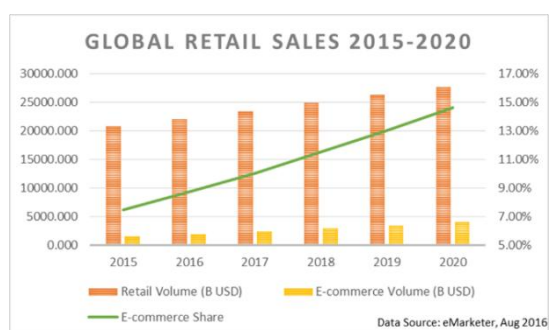

Fig. 1. Growing Significance of E-Commerce in Retail Industry

Along with the growth of e-commerce sector, the count of e-commerce related frauds are also increasing in every year since 1993. As per a report in 2013, 5.65 cents are lost due to frauds out of every $100 in e-commerce turnover. Fraud detection [1] is a key area requiring attention to avoid business losses and to uphold the consumer trust [2] [3]. Most observed frauds in e-commerce industry include stolen credit or debit card information and fraudulent return of products. Over the period of time, researchers have come up with different strategies [4] to detect card related fraudulent actions. The key strategies evolved include Artificial Immune Systems [5], Use of Periodic Features [6], Inductive Learning and Evolutionary Algorithm [7], Hidden Markov Model [8], Neural Data Mining [9], Fusion Approach [10], Bayes Mini-mum Risk Algorithm [11], etc.

Another type of fraud which has become prominent with the evolution of marketplaces is the merchant fraud. These frauds are directly impacting customer satisfaction level and thereby reducing the trustworthiness of the marketplace itself [12] [13]. So marketplace owners are keen in terms of identifying such fraudulent sellers. With the evolution of big data, data mining and machine learning techniques, it is possible to perform analysis on the historic data and correlate it with seller behaviours to identify potential fraudulent moves.

The proposed model results in proactive identification of fraudulent selling attempts in a marketplace with the help of machine learning strategies. Next section of the paper explains the concepts of online marketplace, merchant fraud and also explains the machine learning paradigm. Subsequent sections of this paper deals with the details of the proposed solution frame-work.

## II. ANTECEDENTS

### A. Online Marketplace

An online marketplace [Fig. 2] can be considered as a type of e-commerce portal where products and services are offered by multiple vendors, who may be brands, shops or persons. The marketplace owner will take care of customer attractions and money transactions. The vendors will deal with manufacturing, packaging and shipping. The consumers experience superior shopping experience from online marketplaces and some of the key reasons for the same include

a. More selection opportunities among wide range of products in one place
b. Competitive pricing among sellers
c. Better availability of inventory as multiple vendors sell same product under the same plat-form.
d. Convenience and increased privacy of transactions being done at one place which is not shared with individual sellers.

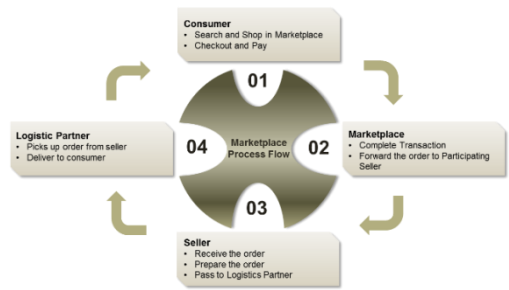

Fig. 2. Depiction of Marketplace Processes

### B. Merchant Fraud

Merchant fraud cover the fraudulent transactions happening from the sellers on marketplace. There are different types of merchant frauds observed. The most prominent one in this category is merchant identify fraud. In this scenario, a criminal impersonate a legitimate seller and charge to consumers' credit or debit card and after collecting some revenue he/she vanishes leaving be-hind all issues like chargebacks and tainted reputation. Another scenario is where some sellers try to create un-due profit by doing some type of malpractice in listing, pricing and shipping of products and services.

### C. Merchant Fraud

Machine Learning [14] is a computer science discipline and a branch of artificial intelligence that deals with computers or machines learning from the past transactions to perform certain tasks and improving its performances with accrued experiences. An intelligent decision making algorithm can be developed using the supervised machine learning technique. This approach requires some existing data, called training set where input scenarios and output scenarios are captured and agreed upon. The learning algorithm do an iterative process to arrive at some logic which will help itself to predict the output for an input scenario where output data is not available [Fig. 3]. The objective of machine learning is to create a model that could make a refined guess in an uncertain scenario in a mechanized manner.

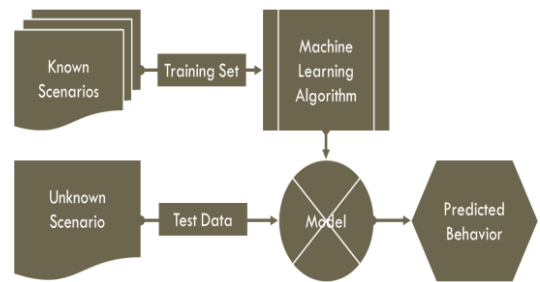

Fig. 3. Supervised Machine Learning Technique

## III. METHODOLOGY

### A. Support Vector Machine

Support Vector Machine (SVM) is a supervised learning model in machine learning space. SVM and associated learning algorithms are powerful tools for classification and regression needs in multiple application domains. Learning from a training set where each candidates are marked into one of the two categories, SVM can build a model which is capable of assigning any new candidate to one of these two categories. This is achieved by algorithmically defining an optimal hyper-plane or decision boundary in the feature space which can function as the formal separator for the classification problem in hand [Fig. 4]. Hence SVM can be used as a non-probabilistic binary linear classifier.

Suppose there is a given data set $(x_1, y_1)$, $(x_2, y_2)$, …, $(x_m, y_m)$, where $y_i = -1$ for inputs $x_i$ in class 0 and $y_i = 1$ for inputs $x_i$ in class 1. Mathematically the classification boundary can be defined as a vector equation for a line in two dimension.

$$\vec{w} \cdot \vec{x} + b = 0 \qquad (1)$$

where $\vec{w}$ and $\vec{x}$ are two dimensional vectors and b is the bias. Further the input vector from class 0 can be defined as the negative support vector $\vec{x}_n$ and the input vector from class 1 as the positive support vector $\vec{x}_p$. Hence the negative classification boundary can be defined as

$$\vec{w}.\vec{x}_n + b = -1 \qquad (2)$$

and the positive classification boundary can be defined as

$$\vec{w}.\vec{x}_p + b = 1 \qquad (3)$$

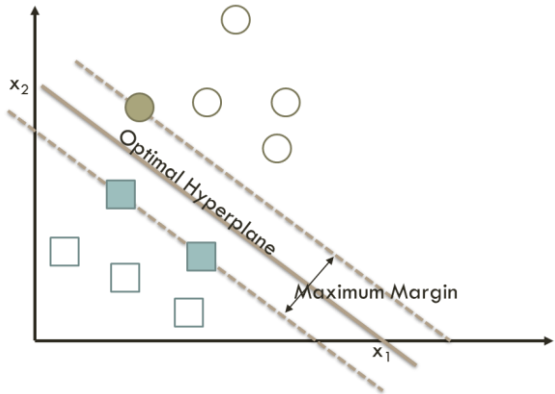

Fig. 4. Maximum-margin Hyperplane of an SVM

So, the distance between the negative and positive classification boundaries is $2/\|\vec{w}\|$ and the size of the margin M is $1/\|\vec{w}\|$. In order to maximize M, it is required to minimize $\vec{w}$. Since all points need to fall to one of the class, following constraints also need to be introduced.

$$\vec{w}.\vec{x}_i + b \leq -1 \text{ for all } \vec{x}_i \text{ in class } 0 \qquad (4)$$

and

$$\vec{w}.\vec{x}_i + b \geq 1 \text{ for all } \vec{x}_i \text{ in class } 1 \qquad (5)$$

So, the SVM can be represented as the following optimization problem

Minimize $\|\vec{w}\|$ subject to
$$y_i(w.x_i - b) \geq 1 \text{ for } i = 1, ..., n \qquad (6)$$

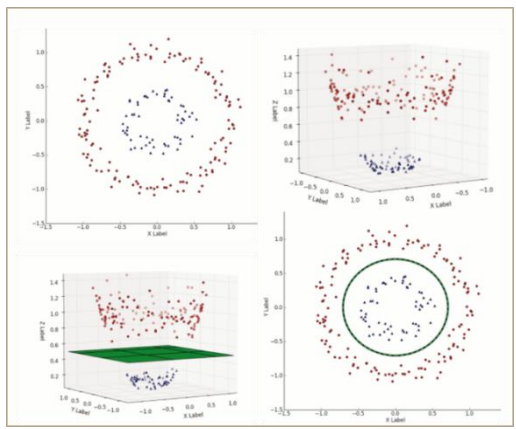

Fig. 5. Kernel Trick: Non-separable dataset at lower dimension is converted to higher dimension to make it separable

SVMs are also capable of doing non-linear classification with the use of kernel trick [Fig. 5] where the inputs are mapped into a high-dimensional feature space to perform classification. The resulting algorithm is similar to that of the linear scenario, except that a nonlinear kernel function will replace every dot product involved. This helps the algorithm to fit the maximum-margin hyper plane in the higher dimensional transformed feature space. Though the classifier identified will be a hyper plane in the transformed feature space, it may get mapped to nonlinear in the original input space.

In general, any mathematical function that confirms to Mercer's condition can be used as a kernel function. However SVMs conventionally use either kernels based on Euclidean distance or kernels based on Euclidean inner products. Kernels are chosen based on the data structure and type of the boundaries between the classes.

SVM is used extensively to decipher a wide range of complex real time scenarios where there is a classification to be made. Hand-written character recognition [15], Image/face detection [16], pedestrian detection [17] and hypertext categorization [18] and fraud detection [19] are some examples of SVM applications. With proper training done to create the model, the problem of over fitting can be reduced considerably in SVM.

## IV. DISCUSSION

This paper proposes a simple framework to perform identification and handling of fraudulent sellers in an online marketplace using machine learning approach, specifically by leveraging the power of SVM for classification [Fig. 6]. This is a multi-stage strategy - gathering all available seller information with the market place, feature extraction from the raw data received, SVM training, performing fraud detection and classification and finally a fraud management module.

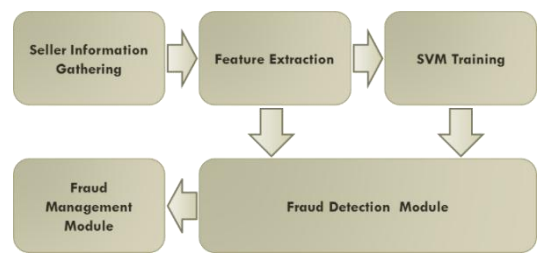

Fig. 6. Fraud Detection Framework

### B. Seller Information Gathering

Marketplaces hold record for every seller activity happening on their platform. This data include product listing details for each seller, transaction details, details regarding order accuracy and timely fulfilment, records of product returns and the causes for the same, customer feedbacks and complaints, etc. These will form critical inputs for feature extraction phase for each of the participating seller in the marketplace.

Another valuable source of information for feature extraction step is inputs from social media analytics. Social media analytics includes the process of gathering, analysing and interpreting social media interactions. Every marketplace keeps an eye on the social media analytics to understand customer sentiments to strategize their marketing programs and to strengthen their customer service activities. Social media analytics can be extended to a further granular level to cover products which can be then attributed to the corresponding sellers in the marketplace.

### C. Feature Extraction

Feature extraction [Fig. 7] deals with the transformation of seller information into set of distinctive features. Features are dissimilar properties that can be observed among input patterns which can help in segregating sellers into different categories in our context. The features considered in the scope of this study includes product listing accuracy, transaction volume, adherence to SLAs set at marketplace level, return ratio, customer satisfaction and/or complaint details. In this scenario considerable volume of input data are available within the marketplace database. Additional information is collected from social media analytics. This feature list form the input criteria for the next step, the SVM training.

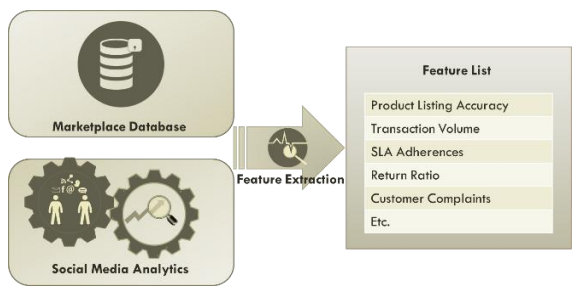

Fig. 7. Feature Extraction from Seller Information

### D. SVM Training

In this approach SVM is used as a non-probabilistic binary linear classifier for sellers on the marketplace. The SVM training algorithm [Fig. 8] learns from the set of training examples where each seller is assigned to one of the two possible categories (normal or fraudulent) to build the classification model. Thus the classification model become capable of classifying any seller into one of the two categories based on the features extracted out of their past traits.

The purpose of an SVM training algorithm is to identify the most optimal hyper plane which will enable it to classify a new candidate passed to it. So training the SVM means finding the solution for a large quadratic programming optimization problem. There are multiple SVM training algorithms available now for consideration. Vapnik who is considered as the co-inventor of SVM suggested an algorithm called chunking algorithm [20]. Osuna et.al. proposed an alternate algorithm which is referred as Decomposition Method [21]. An-other important algorithm is the Sequential Minimal Optimization (SMO) algorithm suggested by Platt [22]. A further refinement on SMO is available in LIBSVM solver (version 2.82) and is proposed by Chang and Lin [23].

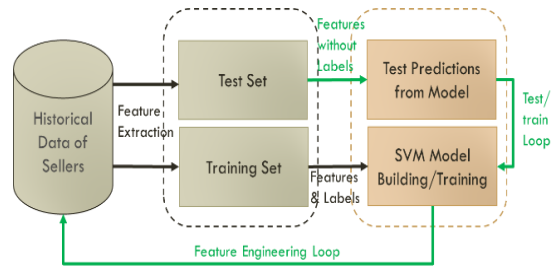

Fig. 8. SVM Training

### E. Fraud Detection Module

Fraud detection module [Fig. 9] is a collection of multiple processes or strategies working independently with their outputs being consolidated to form the final decision on classifying a seller as fraudulent or not. The key constituents of our fraud detection module include
a. Reputation Data
b. Inputs from Fraud Experts
c. Output from Rules Engine and
d. Prediction from SVM Classification

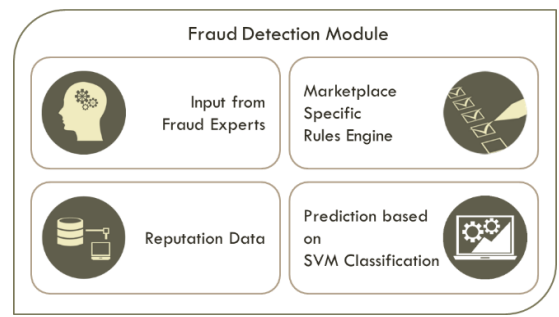

Fig. 9. Fraud Detection Module - Constituents

Reputation data include internal and/or external information pertaining to the reputation of any particular seller. For example, if a seller who is banned for fraudulent action in the past can attempt to re-enrol, may be with a new name. But the fraud detection module can detect such an activity by verifying the key attributes of the seller against its reputation database. Firms make use of external information as well to verify seller reputation like checking their credit score, looking at black listed vendors by their peers, etc.

Further, there are fraud experts working for specific marketplaces or independently. They do research on identifying and exposing new strategies being deployed by frauds on consumer and/or merchant end. Our fraud detection module make use of inputs from such experts to act upon as well as for self-learning which will be used by the SVM classifier. These experts also contribute to the refinement of rules engine as and when they identify a new risk.

Each marketplace maintain their own policies and rules to consider an action as fraudulent or not. The rules engine is a set of checks done on seller actions, attributes or features and are normally implemented as a set of SQL queries. Each rule is verified for compliance and a weighted score is applied. The aggregate score is checked against a defined threshold and the seller will be classified as fraudulent or not. The advantage of a rule engine is that it is easy to add a new rule as soon as a new fraud or risk scenario is detected. However each rule may be short lived, as fraudsters will come up with new tactics as soon as they get caught with one.

SVM is a supervised machine learning technique where the classification model is trained on already known cases of fraud and non-fraud data (called labelled training data set). The training set used as input to the model decide the effectiveness of the SVM classifier. The ability of SVM to predict the label for a new un-labelled data set helps the fraud detection module to cover the scenarios where rule engines are not too effective.

*F. Fraud Management Module*

Fraud management module take the appropriate action on the fraudulent sellers identified by fraud detection module based on the marketplace policies and laws of the land. The actions varies from scenarios to scenario and marketplace to marketplace. In some cases sellers will be given an opportunity to do corrective actions within a stipulated time.

## V. CONCLUSION

For the past few years there is an increased focus on the researches on e-commerce domain, especially on marketplace scenarios mainly attributed to the increasing popularity and year on year growth being exhibited. Reputation is considered as a critical attribute for every online marketplace as consumers has a wide variety of options to choose from. The most important aspect of reputation for an online marketplace is how it protects its customers from fraudulent sellers.

In a normal scenario fraud detection is performed with the help of fraud experts who analyses consumer claims on a product/service being delivered by sellers through the marketplace platform. This can be considered as a reactive measure and most of the time the action on seller ends with an instructions for accepting return of product and/or refund for the customer. This paper proposes a model based on information retrieval and SVM classification to proactively detect fraudulent merchants based on their past performance. In addition the proposed model attempts to leverage the power of social media analytics to understand and con-sider what the society thinks about the services/products delivered by a particular seller through a marketplace.

One key issue which is still open for discussion and hence available for future extensions on this model is the cold start problem. SVM require input data for training the model and hence it is not practically possible to evaluate a new seller as past traits are not available in this case.

## ACKNOWLEDGMENT

I acknowledge and would like to thank Mr. Renjith Ranganadhan for the support, guidance, reviews, valuable suggestions and very useful discussions in the domain of retail marketing.